\documentclass[aps,prl,nobibnotes,twocolumn,superscriptaddress,citeautoscript,showkeys]{revtex4-2}
 \usepackage{ragged2e}
\usepackage{graphicx}
\usepackage{amsmath, amsfonts}
\usepackage{amssymb}
\usepackage[colorlinks=true,linkcolor=blue,citecolor=blue]{hyperref}
\usepackage[utf8]{inputenc}
\usepackage{float}
\usepackage[font=small,labelfont=bf]{caption}

\begin{document}
\emergencystretch 3em

\title{High-speed time-series prediction using compact memristor circuits with adjustable dynamics}

\author{Dániel Molnár}
\affiliation{Department of Physics, Institute of Physics, Budapest University of Technology and Economics, M\H{u}egyetem rkp. 3., H-1111 Budapest, Hungary.\looseness=-1}
\affiliation{HUN-REN-BME Condensed Matter Research Group, M\H{u}egyetem rkp. 3., H-1111 Budapest, Hungary.\looseness=-1}

\author{János Volk Jr.}
\affiliation{Department of Physics, Institute of Physics, Budapest University of Technology and Economics, M\H{u}egyetem rkp. 3., H-1111 Budapest, Hungary.\looseness=-1}

\author{Tímea Nóra Török}
\affiliation{Department of Physics, Institute of Physics, Budapest University of Technology and Economics, M\H{u}egyetem rkp. 3., H-1111 Budapest, Hungary.\looseness=-1}
\affiliation{Institute of Technical Physics and Materials Science,\unpenalty~HUN-REN Centre for Energy Research, Konkoly-Thege M. \'{u}t 29-33, 1121 Budapest, Hungary.\looseness=-1}

\author{Zoltán Balogh}
\affiliation{Department of Physics, Institute of Physics, Budapest University of Technology and Economics, M\H{u}egyetem rkp. 3., H-1111 Budapest, Hungary.\looseness=-1}
\affiliation{HUN-REN-BME Condensed Matter Research Group, M\H{u}egyetem rkp. 3., H-1111 Budapest, Hungary.\looseness=-1}

\author{Nadia Jimenez Olalla}
\affiliation{Institute of Electromagnetic Fields, ETH Zurich, Gloriastrasse 35, 8092 Zurich, Switzerland.\looseness=-1}

\author{Miklós Csontos}
\affiliation{Institute of Electromagnetic Fields, ETH Zurich, Gloriastrasse 35, 8092 Zurich, Switzerland.\looseness=-1}

\author{Juerg Leuthold}
\affiliation{Institute of Electromagnetic Fields, ETH Zurich, Gloriastrasse 35, 8092 Zurich, Switzerland.\looseness=-1}

\author{András Halbritter}\email{halbritter.andras@ttk.bme.hu}
\affiliation{Department of Physics, Institute of Physics, Budapest University of Technology and Economics, M\H{u}egyetem rkp. 3., H-1111 Budapest, Hungary.\looseness=-1}
\affiliation{HUN-REN-BME Condensed Matter Research Group, M\H{u}egyetem rkp. 3., H-1111 Budapest, Hungary.\looseness=-1}


\begin{abstract}
 
Ta$_2$O$_5$ nonvolatile memristors are used as compact, traceable, and well-controllable dynamic reservoir computing layers to perform time-series prediction tasks. The strongly voltage-dependent switching speed is utilized for information processing. It enables the configuration of tailorable programming and forgetting times in response to the positive and negative driving voltage pulses. Benchmarking this framework on time-series prediction problems reveals that the configurable forgetting dynamics enables a high prediction accuracy using a rather small number of memristive input channels. The training is based either on optimizing the output layer using linear regression with fixed forgetting times, or on optimizing the forgetting times as well. In the first case, six memristive channels, while in the second, only two memristive channels are used to demonstrate excellent prediction accuracy for the benchmark tasks. This scheme allows for the tunability of the operating frequency over many orders of magnitude: by adjusting the input voltage levels, the information processing speed of the same memristive dynamic layer can be increased from the kHz to the MHz range while maintaining excellent prediction accuracy. These findings demonstrate the merits of memristor based dynamic networks in the analysis, prediction and recovery of fast temporal signals, approaching telecommunication data rates.
\end{abstract}
\maketitle

Email Address: halbritter.andras@ttk.bme.hu

\justifying


\section{Introduction}

Memristive devices have become central components in modern neuromorphic hardware due to their compact footprint,\cite{Xia2019,Choi2020} low‑energy operation,\cite{Ji2025,Shi2023,Xia2025} and compatibility with CMOS fabrication processes.\cite{Jo2008,Cai2019a,Rao2023} Their ability to store analog conductance states\cite{Rao2023} and to emulate synaptic plasticity\cite{Yang2013b,Yang2015} has enabled a wide range of applications, including signal preprocessing,\cite{Zhao2022,Wang2025} compression,\cite{Wang2023compression,Sun2025} sparse encoding,\cite{Sheridan2017,Ji2019,Woods2019} classification,\cite{Alibart2013,Hu2018,Wang2018c} feature extraction\cite{Choi2017,Shi2025} and combinatorial optimization.\cite{Cai2020, Fehervari2024} In most of these architectures, nonvolatile memristors arranged in crossbar arrays serve as static synaptic weights
demonstrating impressive computational efficiency, particularly in accelerating deep neural network inference at low energy cost.

The reliance on static weight programming, however, introduces substantial overhead during the iterative weight update of the training process.\cite{Jiang2025} Furthermore, solving sufficiently complex problems requires large neural networks that consist of many neural layers. Analyzing signals with temporal correlations also requires memory functionality,\cite{Hochreiter1997} which in turn necessitates the use of recurrent neural networks with feedback connections.\cite{Elman1990} This \emph{network complexity} is illustrated in Fig.~\ref{fig1}a1 with  black/colored arrows for feedforward/feedback connections.
An alternative approach replaces network complexity with \emph{dynamic complexity}.\cite{Jaeger2001} 
In this case, the input data is first pre-processed by a dynamic reservoir layer, whose responses are aggregated into output values by a simple readout layer (see Fig.~\ref{fig1}a2). Most hardware-based (physical) reservoir computing solutions, however, rely on a “black box”-like operation with little insight into the operational dynamics of the reservoir.\cite{Dale2021,Wringe2025,Yan2024}

Memristive devices offer an excellent balance between complexity, traceability and controllability: thanks to their highly voltage-dependent switching speeds, they possess ideal dynamic properties for reservoir computing applications, while the operation of individual elements remains under control.\cite{Du2017,Zhong2021,Cao2022,Csontos2023}. As a result, a simple yet efficient memristive dynamic reservoir layer can be established (see illustration in see Fig.~\ref{fig1}a3). So far mostly volatile memristors exhibiting resistance relaxation have been successfully exploited in reservoir computing applications demonstrating time‑series prediction and dynamical system reconstruction tasks.\cite{Du2017,Liang2022,Midya2019,Milano2022materia,Choi2024,Zhong2021,Kim2026}

In our previous work,\cite{Molnar2025} we demonstrated that 
a single \emph{nonvolatile} Ta$_2$O$_5$ memristor is capable of performing complex temporal signal recognition tasks, like the detection of neural spikes buried in high noise. Furthermore, two Ta$_2$O$_5$ memristors with individually tuned forgetting times are sufficient to learn and predict the evolution of a nonlinear dynamical system with high accuracy. These functionalities rely on the so-called time-voltage dilemma, i.e. the exponential dependence of the switching time on the applied bias (Fig.~\ref{fig1}b). To achieve tunable forgetting using \emph{nonvolatile} memristors, a novel driving method was introduced, where the incoming information is delivered to the system in the form of positive pulses of varying amplitude, while subsequent negative offset pulses enable an adjustable forgetting mechanism (
Figs.~\ref{fig1}c1-c4 and its detailed description later). By adjusting this negative offset level, the memristive channels can be tuned to exhibit longer or shorter term memory operation.

In the present study, we significantly extend this concept. We further exploit the extremely flexible tunability of the operation dynamics. We show that by 
adjusting the input voltage levels, the operating frequency of the very same memristive dynamic layer can be increased from the kHz to the MHz range. This approach leverages the observation that the exponential tunability of the switching time spans many orders of magnitude in the time-domain (Fig.~\ref{fig1}b). This property allows the effective memory time-scale of a memristor to be tuned continuously by adjusting the operating voltage, enabling a single device to encode temporal features across multiple, many orders of magnitude different timescales.  

Moreover, we demonstrate that this broad tunability of the dynamic properties is accompanied by an extremely simple training of the memristive reservoir computing network. We use memristive channels with fixed forgetting times that differ slightly from channel to channel. This is in contrast to the pioneering work that applied identical, physically encoded forgetting times across 90 memristive channels. Furthermore, it also improves our previous approach, where a more complex optimization procedure was required to determine the optimal forgetting times of the two memristive channels. Our current scheme establishes an optimized tradeoff between simple training (linear regression for the output linear combination layer), and a small number of memristive channels. 
We evaluate the performance of our memristive reservoir computing networks on three time-series prediction benchmark tasks of varying complexity, demonstrating that our present scheme provides excellent prediction accuracy using as few as 6 memristive channels.

\section{Results and discussion}
\begin{figure}[t!]
     \includegraphics[width=1\columnwidth]{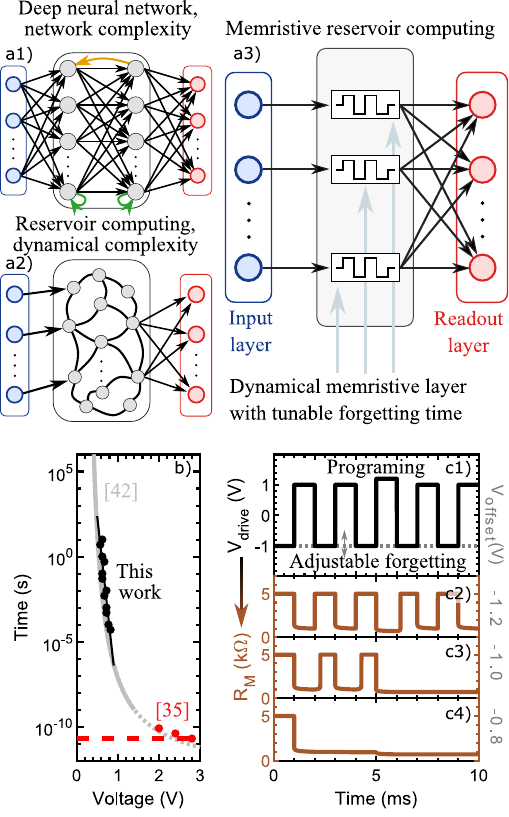}
     \caption{\justifying{\textbf{The concept of our work.} Instead of using complex neural networks with feedback connections (a1), or black-box-like dynamical reservoir computing layers (a2), we apply a compact memristive dynamical reservoir computing layer (a3) to perform time-series prediction tasks. A key feature of our approach is that we can tune the forgetting times of individual memristive channels independently (gray arrows in a3). (b) The tunable forgetting times are enabled by the exponential dependence of the switching time on the applied voltage. The gray curve reproduces a trendline from the literature for Ta$_2$O$_5$ memristors,\cite{Bottger2020} spanning over 15 orders of magnitude in the time-domain, and further extrapolation to even shorter times (gray dashed line). The black dots represent our measurements on the Ta$_2$O$_5$ memristors used in this study. The black line demonstrates a good exponential fit to the data with a single exponent within the investigated time-domain. The red dots demonstrate the results of recent ultrafast time-resolved switching experiments on Ta$_2$O$_5$ memristors, also demonstrating the shortest achieved switching time (dashed horizontal line).\cite{Csontos2023} (c1) A novel driving scheme is applied using positive pulses for programming and an adjustable negative offset for forgetting. Depending on the applied offset level (right labels in c2-c4) the resistance of a single memristive channel exhibits short-term memory response (c2), longer-term memory response due to the middle, higher pulse (c3), or longer-term memory response already due to the first pulse (c3). See text for details.    
     }  }
     \label{fig1}
\end{figure}
In the following, we first present the time-series prediction tasks used to test the predictive performance of our memristive reservoir computing scheme. Next, relying on Ref.~\citenum{Molnar2025}, we illustrate how a single nonvolatile memristor can be used as a complex dynamic system to implement both long-term and short-term memory behavior. We then present the architecture that enables time-series prediction by using multiple memristive channels, each with an individually configurable forgetting time. Afterwards, we demonstrate and analyze the results of our time series prediction analysis.
Finally, we demonstrate how the operation of the memristive time series prediction system can be accelerated by three orders of magnitude -- from the kHz to the MHz range -- by rescaling the drive voltages.

\begin{figure}[t!]
     \includegraphics[width=1\columnwidth]{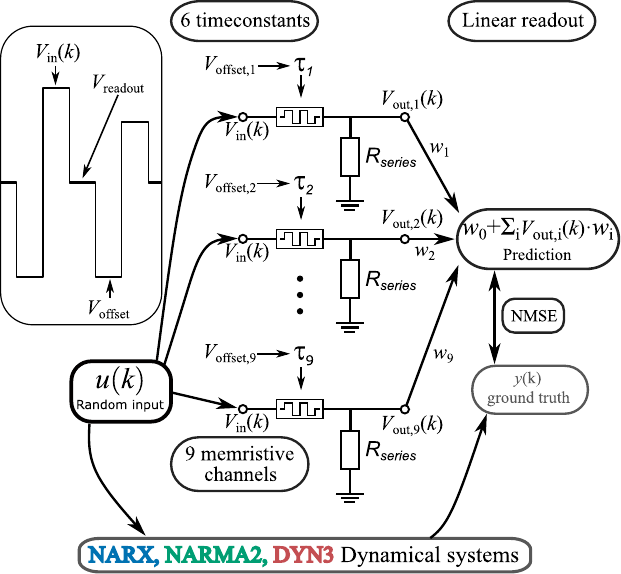}
     \caption{\justifying{\textbf{Illustration of the time series prediction scheme.} Three mathematical dynamical systems (NARX, NARMA2, DYN3) are tested as benchmark tasks, all producing an $y(k)$ output stream in response to a random $u(k)$ input. This random input stream is encoded in the amplitudes of the positive programming pulses ($V_\mathrm{in,i}(k)$), which are fed to the inputs of the various memristive channels, each consisting of one memristor and one resistor in series. Forgetting is achieved by negative offset pulses ($V_\mathrm{offset,i}$), which are independent of $k$, but are different for the different channels ($i$). The dynamical response of the of the memristors is read out as the $V_\mathrm{out,i}(k)$ voltage on the series resistors along the readout pulses. The predicted output ($y_\mathrm{predicted}(k)$) is calculated as the linear combination of the output voltages. The $w_i$ weights of the linear combination are optimized to achieve the best normalized mean squared error (NMSE) between the predicted output and the ground truth ($y(k)$) for the training data, while the prediction accuracy is benchamrked by the NMSE for an independent validation data. In our study 9 memristive channels and 6 different negative offset voltages (different forgetting time-constants) are applied.}}  
     \label{fig2}
\end{figure}

We studied three increasingly complex tasks to benchmark the accuracy of our prediction strategy. All tasks can be described by a equation, where the $k^\mathrm{th}$ state of a mathematical dynamical system $y(k)$, is dependent on an input stream being a uniformly distributed random number $u(k)$ in the range of $[0,0.5]$, and is also dependent on the previous states of the system $y(k-i)$. The different complexity level of the 3 tasks comes from the addition of nonlinearities and adding more previous states to the equations, as listed below:

{
\small
\begin{equation}
   y(k)_\mathrm{NARX}=y(k-1)\cdot(u(k)\cdot0.3+0.2)+0.1
\end{equation}
\begin{equation}
    y(k)_\mathrm{NARMA2}=0.4y(k-1)+0.4y(k-1)y(k-2)+0.6u^3(k)+0.1
\end{equation}
\begin{equation}
y(k)_\mathrm{DYN3}=\frac{y(k-1)\cdot y(k-2)\cdot(y(k-1)+0.25)}{1+y(k-1)^2+y(k-2)^2}+u(k)
\end{equation}
}

The first NARX equation (Nonlinear Autoregressive Exogenous Model)\cite{billings2013nonlinear} is considered as an entry level task, where the output of the dynamical system only depends on the product of its previous state, $y(k-1)$, and an external random input.

The second NARMA2 (Normalized Auto-Regressive Moving Average)\cite{Du2017} equation represents a harder task, where the next state of the system depends not only on its previous state, but also on the state before that, $y(k-2)$
. Furthermore the random input $u(k)$ is cubed introducing an additional nonlinearity to the system. 

The third benchmark task\cite{Atiya2000} does not have a dedicated name in the literature, so we call it Dynamical System 3 (DYN3). This task adds even more nonlinearity by including a fraction of the nonlinear combinations of the previous state and the state before that. 

Next, we illustrate the concept of using single nonvolatile memristors as complex dynamical systems for information processing.\cite{Molnar2025} The key property that enables dynamic complexity is the time-voltage dilemma characteristic of memristors, i.e., the exponential decrease in switching time in response to a linear increase in voltage: 
\begin{equation}
    \tau=10^{-\left(\left|V_{\rm bias}\right|-B\right)/A},
    \label{eq.exp}
\end{equation}
where the $A$ and $B$ parameters may differ for different memristive systems. This type of behavior is demonstrated for our particular Ta$_2$O$_5$ devices by the black measurement points in Fig.~\ref{fig1}b (see the details of voltage-dependent switching time measurements in the ESI). The fitted exponential behavior (black line) is described by $A=0.0415$ and $B=1.42$ parameters. As references, the gray line demonstrates the trend fitted to measurements on similar Ta$_2$O$_5$ devices\cite{Bottger2020} spanning a much broader range (15 orders of magnitude) in the time domain, while the red dots show the results of our previous ultrafast time-resolved studies spanning the $80\,$ps - $15\,$ps time-range.\cite{Csontos2023} The latter, fastest resistive switching is illustrated by the red dashed line. These comparisons underpin the good agreement of our present (black) and previous (red) data with the broader trends reported  in Ref.~\citenum{Bottger2020} (gray solid and dashed lines; the latter representing an extrapolation to ranges beyond the bandwidth). Although the entire time range, covering 17 orders magnitude, cannot be described by a single exponential function—and, in particular, the low- and high-frequency ranges correspond to different A and B parameters—the range relevant to this work (black dots) can be well approximated by Eq.~\ref{eq.exp}. It is to be noted, that the nonvolatile filamentary Ta$_2$O$_5$ memristive devices applied in this work are identical to those in our previous studies, and therefore here only their voltage-dependent switching times are discussed, and further details on the sample preparation and on the resistive switching properties are available in the ESI and in Refs.~\citenum{Molnar2025, Nyary2025}. In the latter two references the same batch of samples were applied as in this work. Further characterization measurements with a special focus on the noise and the ultrafast switching properties on our similarly prepared Ta$_2$O$_5$ thin films and devices are also available in Refs.~\citenum{Csontos2023, Santa2021, Balogh2021}.

Figs.~\ref{fig1}c1 illustrates the driving scheme, which enables tunable forgetting using nonvolatile memristors. The incoming information is programmed into the single memristor and a series connected resistor in the form of positive, variable amplitude driving pulses, while forgetting is achieved by a negative offset level. In the particular example of Fig.~\ref{fig1}c1 the driving sequence includes four identical pulses and a slightly higher pulse in the middle. The resistance response (Figs.~\ref{fig1}c2-c4) of the memristor is simulated according to the exponential voltage-dependence of the switching time (see details in Ref.~\citenum{Molnar2025}) demonstrating a strong dependence on the applied negative offset level. At higher offset, the effect of the programming pulses fades before the arrival of the next pulse (Fig.~\ref{fig1}c2) due to rapid relaxation into the high-resistance state (HRS). At a moderate offset, the middle, higher pulse introduces a longer-lasting stay in the low-resistance state (LRS, see Fig.~\ref{fig1}c3). At an even lower negative offset, already the first programming pulse drives the system into a long-lasting LRS (Fig.~\ref{fig1}c4). Measurements performing the same behavior were reported in our previous work.\cite{Molnar2025} It is to be emphasized that the programming and negative offset voltage levels are shared between the memristor and the resistor in series, $R_\mathrm{series}$. This highlights the important role of the series resistor in the devices' dynamic behavior: as the memristor's resistance approaches the value of the series resistor, the memristor will sense reduced voltages from the driving sequence, which significantly slows down the programming and forgetting speeds. All these rich dynamical features enable nonvolatile memristors to act not merely as static weights but as nonlinear temporal signal processing units, enriching the computational repertoire of neuromorphic circuits.  

In the following, we test our memristive reservoir computing strategy, how accurately it can predict the output values of the previously described 3 dynamical systems. To this end, we use 9 different Ta$_2$O$_5$ nonvolatile memristive devices according to the scheme in Fig.~\ref{fig2}, such that each memristor is considered as separate memristive input chanel. Each memristive channel receives the same $V_\mathrm{in,i}(k)=V_\mathrm{in}(k)=a\cdot u(k) + b$ random input stream such that the $k^\mathrm{th}$ random value is encoded in the amplitude of the positive programming pulse (see Fig.~\ref{fig1}c1). Here, $i$ is the channel number, and the scaling factors are $a=1\,$V and $b=0.5\,$V, which prove to be a suitable operating point for time-series prediction tasks when using 1-ms-long driving pulses.  Inbetween the positive programming pulses, all the memristive channels receive a negative offset voltage, $V_\mathrm{offset, i}$, which is independent of $k$. For all the memristors we use 6 different possible offset levels, decreasing by -0.1~V from -1~V to -1.5~V. In addition to the variable positive programming pulses and the constant negative offset level, our pulsing scheme also includes a readout level of $V_\mathrm{readout}$ after each programming pulse (see top left illustration in Fig.~\ref{fig2}), which is always $0.1\,$V for all the memristive channels. Note, that for the sake of simplicity this part is excluded from the conceptual illustration in Fig.~\ref{fig1}c1, but this readout segment is essential for measuring the state of the memristive channels during periods when the conductance of the memristors remains unchanged. For a given time step $k$, both the programming and the readout, as well as the negative offset pulses, last for $1\,$ms. Each memristor is connected in series with a resistor of $R_\mathrm{series}=1\,k\Omega$. At timestep $k$ the output of each memristive channel is taken as the $V_\mathrm{out,i}(k)$ voltage measured on the series resistor of the given channel during the $k^\mathrm{th}$ readout pulse. Finally, the aggregated output of the whole memristive reservoir computing scheme is calculated as a linear combination of the output voltages, $y_\mathrm{predicted}(k)=w_0+\sum_i V_\mathrm{out,i}(k)\cdot w_\mathrm{i}$. Note, that the readout layer does not include any nonlinear element, i.e. its optimization is a simple linear regression task. To achieve the best prediction accuracy, the response of the memristive layer to a training input, $u_\mathrm{training}(k)$ is measured, and the best-performing $w_i$ weights of the readout layer are determined by linear regression, the latter being calculated in software. Finally, the prediction accuracy is 
quantified through the normalized mean squared error, NMSE$=\frac{\sum_k  (y_{predicted}(k)-y(k))^2}{\sum_k (y(k))^2}$ in response to an independent, validation input data stream, $u_\mathrm{validation}(k)$.

\begin{figure*}[t!]
    \centering
     \includegraphics[width=17cm]{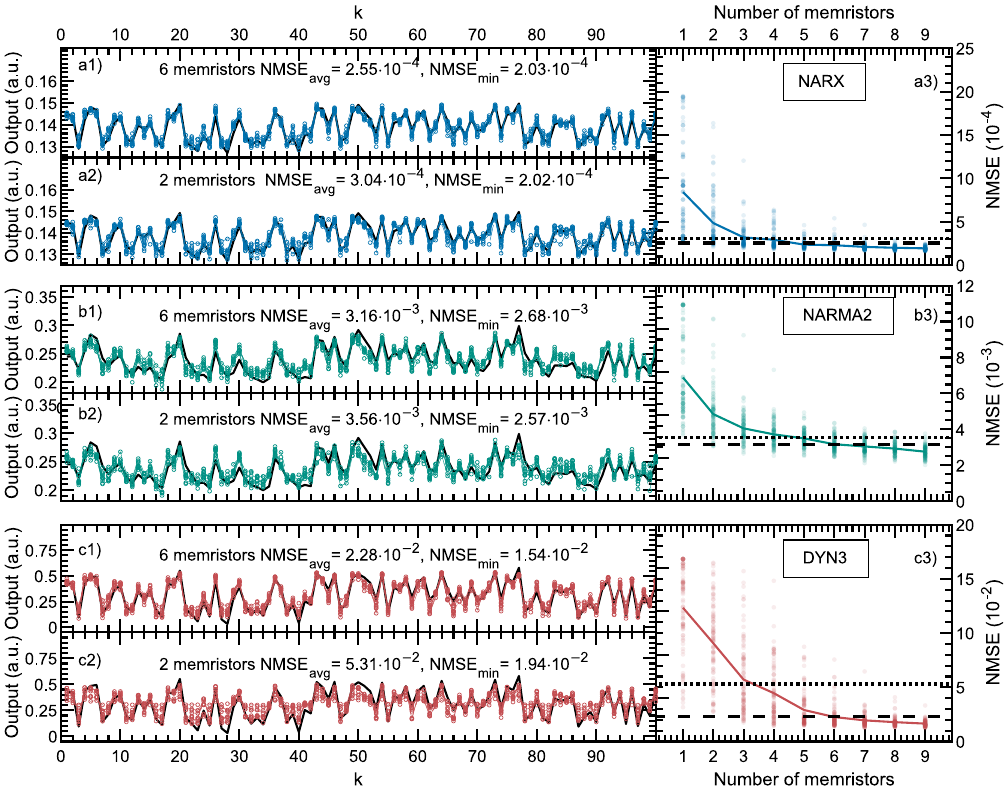}
     \caption{\justifying{\textbf{The results of the time series prediction tasks} using the NARX (a1,a2,a3), the NARMA2 (b1,b2,b3) and the DYN3 (c1,c2,c3) dynamical systems. In Approach I (a1,b1,c1) 6 memristive channels are applied, each using a different (but fixed) negative offset voltage for forgetting (-1\,V, -1.1\,V, -1.2\,V, -1.3\,V -1.4\,V, -1.5\,V). We assign different memristors, selected at random, to the six possible offset voltage levels listed above. The colored curves show the predicted output values for 10 different random selections of the 6 memristors, compared with the ground truth ($y(k)$, black lines). The average and minimum NMSE values are shown above the curves. In this approach solely the $w_i$ weights of the readout linear combination are optimized, and the forgetting times are fixed for the different channels. In Approach 2 (a2,b2,c2) only two, randomly selected memristive channels are used, but beside the $w_i$ weights the offset voltages (forgetting times) of the two channels are also optimized to achieve the best prediction accuracy. The prediction results for the 10 different random memristor selections are compared to the ground truth similarly to Approach I. In Approach III (a3,b3,c3) we calculate the NMSE values as a function of the number of memristive channels (N) such that both the N memristors and the corresponding offset values are randomly selected. For each N the NMSE is calculated for 100 random selections (colored points for a given N). The average NMSE values for Approach III are shown by the solid lines. As a comparison the average NMSE values are also shown for  Approach I (dashed horizontal lines) and Approach II (dotted horizontal lines).}}
     \label{fig3}
\end{figure*}

Next, we investigate the prediction accuracy of our memristive reservoir computing scheme sketched in Fig.~\ref{fig2}. To this end, we apply three different information processing approaches to all three prediction tasks. 

In Approach~I we use 6 memristive input channels ($i=1..6)$, and we apply 6 different offset voltages: $V_{\rm offset,i}=-0.9\,\mathrm{V}-i\cdot 0.1\,\mathrm{V}$. We assign a randomly selected memristor to each offset voltage, meaning we test 10 randomly selected combinations of choosing 6 out of the 9 memristive channels. The results are shown in Figs.~\ref{fig3}a1,b1,c1, where the black lines demonstrate the $y(k)$ mathematical output of the 3 dynamical systems for the validation data sets (NARX in Fig.~\ref{fig3}a1, NARMA2 in Fig.~\ref{fig3}b1 and DYN3 in Fig.~\ref{fig3}c1). This ground truth is compared to the predicted output using 6 memristive channels, as demonstrated by the color lines in Figs.~\ref{fig3}a1,b1,c1, where each figure shows 10 different lines according to the chosen memristor combinations. The minimal variation in the predicted outputs relative to one another, as well as their clear similarity to the ground truth, demonstrates the high accuracy of the predictions
. This finding is further supported by the excellent average and best NMSE values calculated for the validation data (see the NMSE values above the curves in Figs.~\ref{fig3}a1,b1,c1). 
Note that in Approach I all the 6 memristive channels use a fixed offset voltage, which is is slightly different for each channel. This means that during training—that is, when fitting the parameters to a given time-series prediction task—we do not change the offset values; we only adjust the weights of the linear combination used in the readout layer to fit the selected problem. 

As a comparison to the previous approach, next we describe Approach II, where we randomly select 2 memristors, but for these 2 memristors we check the prediction accuracy for all the possible $6\cdot 6$ combinations of the offset values. Finally, we pick the combination, where the best prediction is obtained for the training dataset. The corresponding  predicted outputs are shown by the color lines in Figs.~\ref{fig3}a2,b2,c2 for the validation dataset in comparison to the ground truth (black line). Again, 10 random combinations are used to pick 2 out of the 9 memristive channels, the corresponding 10 predicted outputs are shown by the colored curves plotted on top of each other. The corresponding average and best NMSE values are given above the curves. This approach involves adjusting not only the weights of the linear combination of the output layer during training, but also the negative offset values corresponding to the forgetting time. This is a more complex approach in terms of training than Approach I, but thanks to the optimized forgetting times, fewer memristive channels need to be used and energized during inference, i.e. a more energy-efficient operation with smaller footprint is obtained.

Comparing Approach I and II we can generally say, that the average NMSE of Approach I is slightly below to that of Approach II for the NARX and NARMA2 problem, while Approach I gives a significantly better solution to DYN3 than Approach II. In this more complex DYN3 prediction task, it is already evident from a comparison of the curves in Fig.~\ref{fig3}c1 (Approach I) and Fig.~\ref{fig3}c2 (Approach II) that the latter shows severe deviations from the true values (black line), while the former provides a substantially better prediction. This means that in Approach II, optimizing the forgetting times is not sufficient for an accurate prediction; that is, Approach I, which uses 6 memristors, not only simplifies training but the larger number of memristive channels is is fundamentally essential for an accurate prediction.

The comparison of Approach I and II  is especially interesting for the for the NARMA 2 problem, where two previous benchmark results are available from the literature.\cite{Du2017,Molnar2025} In the first paper\cite{Du2017} 90 volatile memristors were applied for the time series prediction tasks achieving NMSE$=3.13\cdot10^{-3}$. In that case  the forgetting times were non-tunable, i.e. different effective response times were achieved by the temporal rescaling of the input stream for the various channels. The second paper\cite{Molnar2025} represents our previous work, where only 2 memristive channels with well-tunable time-constants were applied (NMSE$=3.03\cdot10^{-3}$). As an important conclusion, we can state that our approach, which uses 6 memristive channels with 6 fixed, different offset voltages, provides slightly better prediction accuracy for NARMA2 than the two previous approaches, which represent the two extremes (many channels with non-adjustable forgetting times, and only 2 channels with fully optimized forgetting times).

Finally, we discuss Approach III, where we study the prediction accuracy as a function of the $N$ number of memristive channels applied, and furthermore, we do not apply different offset values for each channel, rather the offset values are randomly selected in the $-1.. -1.5\,$V range, i.e. different channels may have the same offset. In particular, for each N we test 100 different combinations, such that one combination includes $N$ randomly selected different memristors, and a random choice of the offset values for each memristive channel. By using this random selection, it is possible to test the extent to which prediction accuracy depends on the precise setting of dynamic parameters, and on the choice of the actual devices. The results are summarized in Figs.~\ref{fig3}a3,b3,c3 by the colored dots. Even this random approach delivers rather good prediction accuracies once a sufficiently large number of memristive channels are used. For a quantitative comparison, the solid lines in Figs.~\ref{fig3}a3,b3,c3 show the average NMSE values for Approach III as a function of N, while the reference average NMSE values taken from Approach I/Approach II are shown by the horizontal dashed lines/dotted lines. For N=6, the random offset selection yields results similar to those of Approach I (6 channels with different offsets) for all tasks; while with a further increase in N, even better prediction accuracy than that of Approach I can be achieved.

\label{2p3}

 \begin{figure}[t!]
     \includegraphics[width=1\columnwidth]{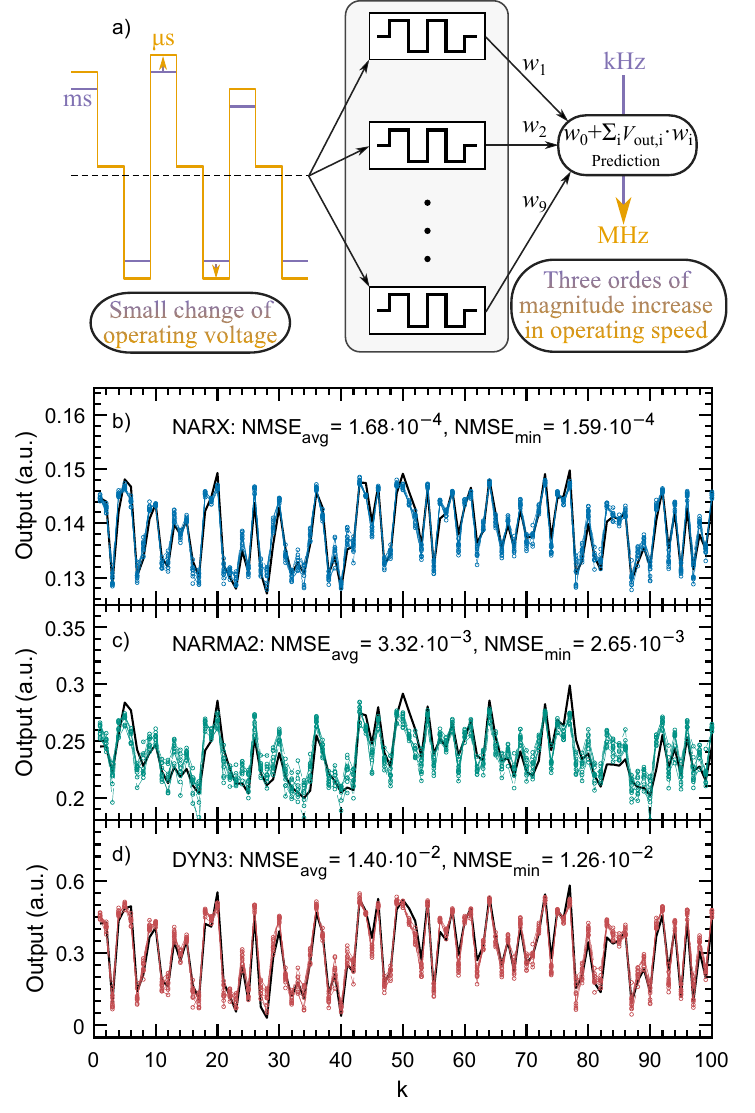}
     \caption{\justifying{\textbf{Memristive time series prediction at MHz clock frequency.} (a) The scheme of the frequency-speed-up. Whereas in Fig.~\ref{fig3} the programming, offset and readout pulses were $1\,$ms long (see the purple input stream on the left side), here $1\,\mu$s long pulse segments are applied (orange input stream on the left side). To speed up the programming and forgetting accordingly, the amplitudes of the programming and offset pulses are enhanced (see the difference between the orange and purple input streams). (b,c,d) The results of the time series prediction for the NARX (b), NARMA2 (c) and DYN3 (d) tasks using Approach II and $1\,\mu$s long driving pulse segments. The colored curves show the predicted output values for 10 different random selections of the 6 memristors, compared with the ground truth ($y(k)$, black lines). The average and minimum NMSE values are shown above the curves. The results for Approach II and III are available in the ESI.}}
     \label{fig4}
\end{figure}

So far we have tested our memristive time-series prediction scheme at kHz clock frequency, i.e. using programming, readout and offset pulses with $1\,$ms lengths. However, the black line in Fig.~\ref{fig1}b demonstrates that a small change of the operating voltage yields orders of magnitude faster switching times. Taking advantage of this property, we are now testing whether we can increase the speed of time-series prediction by three orders of magnitude, from the kHz range to the MHz range. To this end, we apply three orders of magnitude faster input signal, where the $1\,$ms long programming, readout and offset pulses (see purple pulse scheme in Fig.~\ref{fig4}a) are replaced by $1\,\mu$s long ones (orange pulse scheme in Fig.~\ref{fig4}a). To speed up the programming and forgetting of the memristive channels, the amplitude of the programming and offset pulses are increased by $0.2\,$V (see the difference between the orange and purple curves in Fig.~\ref{fig4}a). This faster operation required a somewhat modified experimental setup, including a high-frequency current amplifier (see the ESI for the details). With this high-frequency driving scheme we measured the response of the 9 memristive channels at the 6 different offset values in the $[-1.2,-1.7]$~V range, optimized the readout layer's weights for the the best prediction accuracy, and evaluated the NMSE values for the validation dataset. The results of these studies are shown in Figs.~\ref{fig4}b,c,d for Approach I, while the results for Approach II and III are provided in the ESI. Figs.~\ref{fig4}b,c,d demonstrate that the prediction accuracy is even somewhat better compared to the case where the same problem was treated at 3 orders of magnitude lower frequency (Figs.~\ref{fig3}a1,b1,c1). This means that by correctly selecting the operating voltage levels of the programming and the offset pulses, we can tune the dynamic properties of the memristive reservoir computing layer to the typical timescales of the problem to be solved, thereby allowing us to very easily change the operating frequency range—even by orders of magnitude. 

Finally, we note that, according to the red dots in Fig.~\ref{fig1}b and Ref.~\citenum{Csontos2023} the device physics would enable even higher frequency operation, reaching frequencies above GHz. However, fully exploiting this potential requires either a sophisticated, impedance-matched measurement setup or, preferably, a higher level of integration, with all the necessary circuitry implemented on a single chip.

\section{Conclusions}

In conclusion, we presented a memristive reservoir computing scheme suitable for effective time series forecasting. The input data are processed by a dynamic reservoir layer consisting of nonvolatile memristors. The predicted time series are generated from the outputs of this layer using a simple linear combination. To use nonvolatile memristors as a dynamic reservoir suitable for tunable forgetting, we program the incoming information into the system in the form of positive pulses. Forgetting is achieved with a negative offset. By exploiting the time-voltage dilemma characteristic of memristive systems, we show that the dynamics of memristive channels can be tailored in our scheme. By rescaling the voltage range of the pulses that trigger programming and forgetting, we can accelerate the operation by several orders of magnitude, from the kHz range to the MHz range. In addition to the order-of-magnitude tuning, we can also fine-tune the dynamic properties of individual memristive channels, by setting different forgetting times for each channel. When tested on benchmark time series forecasting tasks, our memristive time series forecasting system demonstrates excellent forecasting accuracy at both kHz and MHz operating frequencies.

We compared our memristive time-series prediction system with the two extreme cases available in the literature for the NARMA 2 prediction task. In the first, a large number of memristive channels (90) were used with fixed, hardware-encoded forgetting times,\cite{Du2017} while in the second (our previous work\cite{Molnar2025}), a minimal number of memristive channels (only two memristors) were used, but the forgetting times were optimized individually during training. Here, we demonstrated an ideal trade-off between these two extreme cases. On one hand, we preset the forgetting times for each channel, thereby simplifying the training to a linear regression on the weights of the output layer. On the other hand, thanks to the differently configured forgetting times for each channel, a surprisingly small number (only 6 memristive channels) is sufficient to achieve efficient time series prediction. 

With this, we have demonstrated that nonvolatile memristors can not only be used as static weights in hardware accelerators for neural networks, but are also ideally suited for implementing compact, traceable, well-controllable, and efficient dynamic reservoir computing layers. Thanks to the channel-specific forgetting times, the easy scalability of the number of channels, and the flexible tuning of the operating frequencies, this provides a versatile method for the efficient analysis and prediction of signals with temporal correlations.
The MHz frequencies achieved, along with the promise of even faster performance resulting from the devices’ ultra-fast operation, pave the way for applications at telecommunications frequencies, where time series analysis, prediction and recovery tasks are highly relevant.

\section*{Supporting information}
Supporting Information is available from the authors.

\section*{Data availability}
The data that support the findings of this study are available at the authors upon reasonable request.

\section*{Code availability}
The core computer code of our simulations is available at the authors upon reasonable request.

\section*{Acknowledgements}

This research was supported by the NKFI K143169, K143282, and 152611 grants. T.N.T. acknowledges the support of the Bolyai J\'{a}nos Research Scholarship of the Hungarian Academy of Sciences. J.L., M.C. and N.J.O. acknowledge the financial support of the Werner Siemens Stiftung. Z.B. acknowledges the support of the NKFI FK146339 grant.

\section*{Author contributions}

The measurements and the data analysis were performed by D.M. with contribution of J.V.Jr. in the measurement of the time-voltage dilemma. The Ta$_{2}$O$_{5}$ memristors were developed and fabricated by M.C and N.J.O. in the group of J.L. T.N.T. and Z.B. respectively contributed to the optimization of the devices and the measurement setup. The project was conceived and supervised by A.H. The manuscript was written by D.M. and A.H. All authors contributed to the discussion of the results.

\section*{Conflict of interest}
The authors declare no conﬂict of interest.

\bibliographystyle{MSP}
\bibliography{References.bib}

\clearpage
\newpage

\end{document}


\title{High-speed time-series prediction using compact memristor circuits with adjustable dynamics\\ \vspace{0.5cm}
Supporting Information} \vspace{0.2cm}

\author{Dániel Molnár}
\affiliation{Department of Physics, Institute of Physics, Budapest University of Technology and Economics, M\H{u}egyetem rkp. 3., H-1111 Budapest, Hungary.\looseness=-1}
\affiliation{HUN-REN-BME Condensed Matter Research Group, M\H{u}egyetem rkp. 3., H-1111 Budapest, Hungary.\looseness=-1}

\author{János Volk Jr.}
\affiliation{Department of Physics, Institute of Physics, Budapest University of Technology and Economics, M\H{u}egyetem rkp. 3., H-1111 Budapest, Hungary.\looseness=-1}

\author{Tímea Nóra Török}
\affiliation{Department of Physics, Institute of Physics, Budapest University of Technology and Economics, M\H{u}egyetem rkp. 3., H-1111 Budapest, Hungary.\looseness=-1}
\affiliation{Institute of Technical Physics and Materials Science,\unpenalty~HUN-REN Centre for Energy Research, Konkoly-Thege M. \'{u}t 29-33, 1121 Budapest, Hungary.\looseness=-1}

\author{Zoltán Balogh}
\affiliation{Department of Physics, Institute of Physics, Budapest University of Technology and Economics, M\H{u}egyetem rkp. 3., H-1111 Budapest, Hungary.\looseness=-1}
\affiliation{HUN-REN-BME Condensed Matter Research Group, M\H{u}egyetem rkp. 3., H-1111 Budapest, Hungary.\looseness=-1}

\author{Nadia Jimenez Olalla}
\affiliation{Institute of Electromagnetic Fields, ETH Zurich, Gloriastrasse 35, 8092 Zurich, Switzerland.\looseness=-1}

\author{Miklós Csontos}
\affiliation{Institute of Electromagnetic Fields, ETH Zurich, Gloriastrasse 35, 8092 Zurich, Switzerland.\looseness=-1}

\author{Juerg Leuthold}
\affiliation{Institute of Electromagnetic Fields, ETH Zurich, Gloriastrasse 35, 8092 Zurich, Switzerland.\looseness=-1}

\author{András Halbritter}\email{halbritter.andras@ttk.bme.hu}
\affiliation{Department of Physics, Institute of Physics, Budapest University of Technology and Economics, M\H{u}egyetem rkp. 3., H-1111 Budapest, Hungary.\looseness=-1}
\affiliation{HUN-REN-BME Condensed Matter Research Group, M\H{u}egyetem rkp. 3., H-1111 Budapest, Hungary.\looseness=-1}

\maketitle
\thispagestyle{fancy}
\lhead[]{Supporting Information}
\rhead[]{}

\newpage

\setcounter{secnumdepth}{1}

\section{Sample fabrication}

The Ta/Ta$_2$O$_5$/Pt memristors were fabricated on Si/SiO$_2$ substrates with a 280~nm thick thermally grown SiO$_2$ layer. First, a 10~nm thick Ti adhesion layer and a 40~nm thick Pt bottom electrode were deposited by electron-beam evaporation at a base pressure of $10^{-7}$~mbar and a deposition rate of 0.1~nm/s. Subsequently, a 5~nm thick Ta$_2$O$_5$ switching layer was deposited by reactive high-power impulse magnetron sputtering (HiPIMS) from a Ta target at a pressure of 6~mTorr using Ar and O$_2$ flow rates of 45 and 5~sccm, respectively, and an RF power of 250~W. Finally, the Ta top electrode and Pt capping layer were deposited by sputtering at 4~mTorr using an Ar flow of 45~sccm, with RF and DC powers of 250~W and 125~W for Ta and Pt, respectively. Both the 2.5~$\mu$m wide bottom and top electrodes were defined by standard optical lithography followed by lift-off.

\section{Measurement Methods}

\begin{figure}[b!]
\includegraphics[width=0.5\columnwidth]{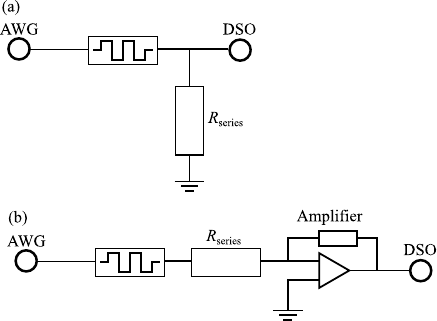}
\caption{{The measurement setup} (a) used for the kHz time series prediction, while the (b) is used for the the MHz operation of time series prediction and the measurement time-voltage dilemma}
     \label{fig_s1}
\end{figure}

Figure~\ref{fig_s1}a,b demonstrates the two schemes we have applied in our measurements. In both cases the memristive channels are driven by an Arbitrary Waveform Generator (AWG). Panel (a) demonstrates an approach, where the $V_\mathrm{out,i}$ voltage on the $R_\mathrm{series}$ series resistor of channel $i$ (see Fig.~2 in the main text) is directly measured by a Digital Sampling Oscilloscope (DSO). This approach is appropriate for the low frequency measurements, however, in the MHz range an impedance matched circuit is more appropriate. Using the scheme of Fig.~\ref{fig_s1}a unwanted reflections will occur at the high impedance (High Z) input of the DSO, while a $50\,\Omega$ input of the DSO would short-circuit the resistor in series. Therefore, Fig.~\ref{fig_s1}b is a better suited  circuit for high frequency measurements, where instead of directly measuring $V_\mathrm{out,i}$, rather the $I_i$ current of channel i is measured with a transimpedance amplifier. The latter has $50\,\Omega$ impedance matched input and output, and the output is  connected to the $50\,\Omega$ input of the DSO.

Time-series prediction experiments at kHz frequencies were performed using the measurement setup shown in Fig.~\ref{fig_s1}(a). The input signals were generated by a National Instruments PXIe-5433 arbitrary waveform generator (AWG) with 80~MHz analog bandwidth, while the voltage drop across the series resistor was recorded using the High-Z (1~M$\Omega$) input of a PicoScope PS6404A oscilloscope with 500~MHz analog bandwidth. Along the readout pulses the transient regions were excluded from the data, and the remaining part ($\approx 1000$ datapoints) was averaged to obtain $V_\mathrm{out,i}$. The outputs of the various memristive channels were measured separately, and the predicted output was then obtained by combining the measured signals in software.

To extend the operating frequency into the MHz regime, the modified setup shown in Fig.~\ref{fig_s1}b was employed. The same AWG and PicoScope PS6404A oscilloscope were used, with the oscilloscope configured for 50~$\Omega$ input termination to ensure proper impedance matching, while a Femto DHPCA amplifier with 200~MHz analog bandwidth provided the high-frequency readout required for the MHz measurements. In this case the output voltage was evaluated as $V_\mathrm{out,i}=R_\mathrm{series}\cdot I_i$.
Apart from these hardware modifications, the experimental protocol remained unchanged. The same programming, readout, and data-processing procedures were therefore used for both frequency ranges. 

For the characterization of the time-voltage dilemma(see the black datapoints in Fig.~1b of the main text) also the measurement setup in Fig.~\ref{fig_s1}b was used. Voltage waveforms were generated using a Zurich Instruments HDAWG arbitrary waveform generator (750~MHz analog bandwidth). For signals below $500~\mathrm{kHz}$, the output was amplified using a Femto DLPCA-200 amplifier, while signals above $500~\mathrm{kHz}$ were amplified using a Mini Circuts ZHL-72A+ 700~MHz amplifier. The voltage waveforms were monitored at the 50~$\Omega$ input of a PicoScope PS6404A oscilloscope.

Prior to each measurement, the memristor was initialized to a resistance of $R_\mathrm{initial}=20~\mathrm{k}\Omega\pm 500\,\Omega$ using an iterative programming procedure. Starting from the low-resistance state, negative voltage pulses with an initial amplitude of $-500~\mathrm{mV}$ were applied. If the measured resistance increased, pulses of the same amplitude were repeated until the target resistance was reached. Otherwise, the pulse amplitude was increased in steps of $50~\mathrm{mV}$. If the resistance exceeded the target value, the same procedure was repeated using voltage pulses of opposite polarity until the device was restored to the desired resistance region. 

After initialization, a programming pulse with amplitude $V_\mathrm{set}$ and duration $t_\mathrm{set}$ was applied, followed by a $100~\mathrm{mV}$ read pulse to determine the resulting resistance. Pulse durations ranging from $10~\mathrm{s}$ to $5\times10^{-5}~\mathrm{s}$ were applied covering altogether $12$ different pulse widths distributed evenly along the logarithmic time axis. At a certain pulse width, first pulses with lower (non-switching) amplitudes were applied, and the pulse amplitude was gradually increased in $50\,$mV steps. The response to each pulse was measured, and after each pulse $R_\mathrm{initial}$ was restored, if a significant switching was observed. The switching voltage at a certain pulse width is taken as the smallest voltage amplitude, where the resistance decreases below the half of the initial resistance. The obtained switching time/voltage pairs are displayed on Fig~1b of the main text.

\section{{Time series prediction at MHz frequencies}}

\begin{figure}[t!]
\includegraphics[width=\columnwidth]{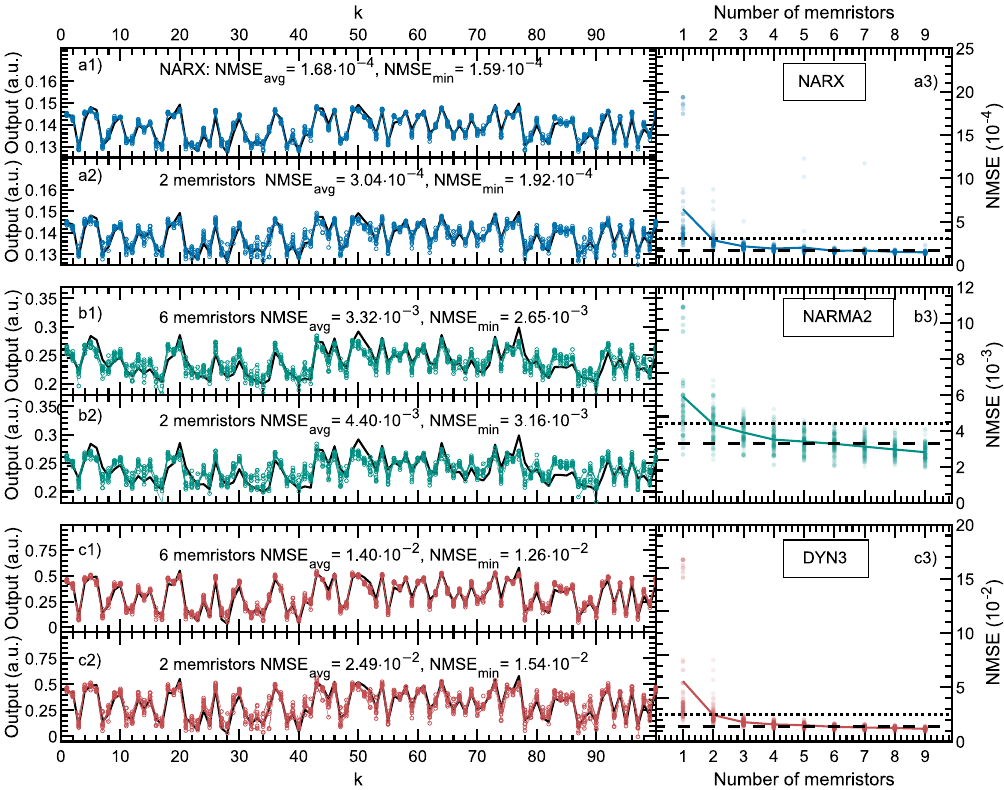}
\caption{{The results of the time series prediction tasks} using the NARX (a1,a2,a3), the NARMA2 (b1,b2,b3) and the DYN3 (c1,c2,c3) dynamical systems in the MHz operating range. In Approach I (a1,b1,c1) 6 memristive channels are applied, each using a different (but fixed) negative offset voltage for forgetting (-1.2\,V, -1.3\,V, -1.4\,V, -1.5\,V -1.6\,V, -1.7\,V). We assign different memristors, selected at random, to the six possible offset voltage levels listed above. The colored curves show the predicted output values for 10 different random selections of the 6 memristors, compared with the ground truth ($y(k)$, black lines). The average and minimum NMSE values are shown above the curves. In this approach solely the $w_i$ weights of the readout linear combination are optimized, and the forgetting times are fixed for the different channels. In Approach 2 (a2,b2,c2) only two, randomly selected memristive channels are used, but beside the $w_i$ weights the offset voltages (forgetting times) of the two channels are also optimized to achieve the best prediction accuracy. The prediction results for the 10 different random memristor selections are compared to the ground truth similarly to Approach I. In Approach III (a3,b3,c3) we calculate the NMSE values as a function of the number of memristive channels (N) such that both the N memristors and the corresponding offset values are randomly selected. For each N the NMSE is calculated for 100 random selections (colored points for a given N). The average NMSE values for Approach III are shown by the solid lines. As a comparison the average NMSE values are also shown for  Approach I (dashed horizontal lines) and Approach II (dotted horizontal lines).}
     \label{fig_s2}
\end{figure}

In the main text (Fig.~4) the high (MHz) frequency time series prediction is only demonstrated for Approach I. Here, Fig.~\ref{fig_s2} demonstrates the results of the same measurements evaluated for Approach I, II and III as well,  following the visualization scheme of the low frequency measurements (Fig.~3 in the main texts). The detailed description of these approaches is available in the main text and it is also summarized  in the caption of Fig.~\ref{fig_s2}.

